\documentclass[intlimits,twoside,a4paper]{article}

\usepackage{amsmath,amssymb}
\usepackage{graphicx}

\usepackage[T2A]{fontenc}
\usepackage[cp1251]{inputenc}

\usepackage[eqsecnum]{cmpj3}

\issue{2019}{22}{3}{33701}
\doinumber{10.5488/CMP.22.33701}

\title[Influence of  nanoparticle surface and shape on the 
dipole magnetic absorption]{Influence of  nanoparticle surface and shape on 
the dipole magnetic absorption of ultrashort laser pulses
}
\author[N.I. Grigorchuk]{N.I. Grigorchuk}
\address{Bogolyubov Institute for Theoretical Physics of the National Academy of Sciences of Ukraine, \\
 14-b Metrologichna St., 03143 Kyiv, Ukraine
}

\date{Received June 13, 2019, in final form July 9, 2019}

\DeclareMathOperator{\rot}{rot}

\begin{document}

\maketitle

\begin{abstract}
The theory on the magnetic field energy absorption by metal nanoparticles
of a nonspherical shape irradiated with ultrashort laser pulses of different
duration is developed. The effect of both the particle surface and the
particle shape on the absorbed energy is studied. For the particles
having an oblate or prolate spheroidal shape, the dependence of this
energy on the orientation of the magnetic field upon a particle, the
degree of its deviation from a spherical shape, a pulse duration, and
the carrier frequency of the laser ray are found. A significant
increase in the absorption is established when an electron mean free
path coincides with the size of the particle. The Drude and kinetic
approaches are used and the results are compared with each other.
\keywords metal nanoparticles, dipole magnetic absorption,
ultrashort laser pulses, nonspherical particles

\pacs 78.67.-n, 65.80.-g, 73.23.-b, 68.49.Jk, 52.25.Os
\end{abstract}

\section{Introduction}
\label{intro}
With the help of pulses of short duration, it is possible to study
the dynamics of rapid processes occurring in atoms, molecules and
solids. Pico- and femtosecond resolution allows one to study oscillatory
and rotational intra-molecular movements, carrier dynamics in
semiconductor nanostructures, phase transitions in solids, formation
and breakdown of chemical bonds, etc. \cite{AVC,CJS}.

In recent years, there has been a continuous experimental attention
paid to the study of ultrashort dynamics of electrons in metallic
nanoparticles (MNs). The increase of local magnetic fields close to
MNs makes them useful as markers in biological systems \cite{BTM,AHL},
in biosensing for magnetic particle detection techniques \cite{CKZ},
as well as in magnetic diagnostic systems \cite{LSC}.
In general, nanostructures are widely used in modern
high-speed electronics and optoelectronics.

Modern optical devices provide an opportunity to record the optical
response of a single nanoparticle and thus allow one to study the
properties of separate nanobjects \cite{ACF}. This opens up
new direct possibilities for sensing multi-electronic dynamics
in limited systems.

Theoretical treatment of the magnetic dipole absorption was
proposed by Wilkinson and coauthors in \cite{WMW}. The
enhanced absorption of a surface plasma wave by MNs in the presence
of an external magnetic field was studied recently in \cite{DCV}.
The possibility for nanoparticles to produce a strong magnetic dipole
absorption at optical frequencies has also been noticed by authors of
\cite{AMM}. The major role of magnetic field losses in microwave
heating of metal was demonstrated in \cite{CRA}.

Interesting results on the aforementioned problems were recently obtained in
\cite{ENZ,KMB,GCV,KCZ,G}. Particularly, in \cite{KMB,ENZ},
it was demonstrated that the magnetic dipole resonances and the magnetic
response can be detected even in the dielectric nanospheres.

As is known, the absorption by MNs in the field of a
monochromatic electromagnetic (EM) wave, whose length is much larger
than the size of the particle, is due to the contribution both from the
electrical component of the EM wave (electrical absorption) and the
magnetic component (magnetic absorption) \cite{BH,ASS}.

 In previous studies \cite{G1,TG,GT1,GT2}, 
 we have shown that depending on the size of the particle,
 as well as on the frequency and polarization of the wave, the magnetic
 absorption can be either larger or smaller than the electric one.

 The situation may change with the use of ultrashort pulses. Firstly,
 the ultrashort pulse contains almost all the harmonics, including
 those that coincide with plasmon resonances.
It allows all the resonances inherent in the system to manifest themselves,
and to study the system response in general. Secondly, the frequency
distribution over the ultrashort pulse spectrum is highly heterogeneous
(Gaussian). The above factors may change the relation between the
 electrical and magnetic absorption contribution.

We studied the features of electrical absorption of ultrashort pulses in \cite{GT1}.
In this paper, we focus on the research of the absorption characteristics
due to the effect of the magnetic component of the ultrashort EM pulse.
This problem remains little studied, especially for the MS of a
non-spherical form with dimensions both smaller and larger than the
electron mean free path. The main goal of this study is to investigate
how the surface and shape of the particle can manifest themselves
in the dipole magnetic absorption of ultrashort laser pulses.

The work is structured as follows. The second section describes
the model and the main principles of the problem. The third
section is devoted to the study of magnetic absorption with
Drude and kinetic approaches. The fourth section discusses
the results and the fifth section presents the main results
and conclusions.

\section{Model and main principles}
\label{sec:1}

Let the laser pulse fall on the MN, the electric field
of which we take in the form \cite{GT1}
\begin{equation}
\label{eq1}
{\rm {\bf E}}({\rm {\bf r}},t) = {\rm {\bf E}}_{0} \re^{ - \Gamma ^{2}
\left(t - \frac{ {\rm\bf k}_\text{c} {\rm\bf r}}{\omega_\text{c}}
\right)^{2}}\cos {\left[ {\omega_\text{c} \left( {t -
{\frac{{{\rm {\bf k}}_{\text{c}} {\rm {\bf r}}}}{{\omega _\text{c}}} }} \right)} \right]},
\end{equation}
where $\Gamma $ is the value reversed to the duration of the laser pulse,
$\omega_\text{c} $ is a carrier frequency of the EM wave, $\vert {\rm {\bf k}}_{\text{c}} \vert \; = \omega_\text{c} / c$, and ${\rm {\bf E}}_{0} $ has the sense
of the maximum value of the electric field in the pulse. In addition to
the electric field, the laser pulse field also has a magnetic component,
which is connected with the electrical component by the corresponding
Maxwell equation
\begin{equation}
 \label{eq2}
  \rot{ {\bf E}}({\rm {\bf r}},t) = - {\frac{{1}}{{c}}}{\frac{{
   \partial}}{{\partial t}}}{\rm {\bf H}}({\rm {\bf r}},t).
\end{equation}

By taking ${\rm {\bf E}}({\rm {\bf r}},t)$ in the form of equation~(\ref{eq1}),
we establish with the help of equation~(\ref{eq2}), the value of
the magnetic field. Most simply, this connection will be seen
between the Fourier components of these quantities.

Making the Fourier transform in equation~(\ref{eq2}), and performing the
integration over time in the limits $(-\infty, \infty)$, we find
for the value of the magnetic field the form
\begin{equation}
\label{eq3}
{\rm {\bf H}}({\rm {\bf r}},\omega ) = \frac{\sqrt {\piup} }
{2\Gamma}\left[ \re^{ -{\frac{(\omega - \omega_\text{c} )^2}{4\Gamma^2}}} +
\re^{ -{\frac{(\omega + \omega_\text{c} )^2}{4\Gamma ^2}}} \right]
\re^{\ri{\rm\bf k}_\text{c} {\rm\bf r}\big( \frac{\omega} {\omega_\text{c} } \big)}
\rm{\bf m} \times {\rm {\bf E}}_0\,,
\end{equation}
where $ {\rm\bf m} =  {\rm\bf k}_\text{c} / k_\text{c} $ is
the unit vector directed along the direction of the EM wave
propagation.

The magnetic field of the laser pulse generates 
the \textit{rotational} electric field ${\rm {\bf E}}_\text{rot} ({\rm {\bf r}},t)$ inside the MN.
To find an expression for the energy absorbed by MN, it is necessary
to know the internal field ${\rm {\bf E}}_\text{rot} ({\rm\bf r},t)$.
In order to determine it, we pay attention to the multiplier
feature of the Fourier component in equation~(\ref{eq3})
related to the coordinate dependence.
If the characteristic size of the nanoparticle $R$ is such that
$k_\text{c} R \ll 1$, that is, the length of the carrier wave is much
greater than the size of the MN, then the coordinate dependence
of the Fourier-component ${\rm{\bf H}}({\rm {\bf r}},\omega )$
within the particle can be neglected. This means that it is
possible to determine the Fourier component of the internal
field inside the MN as a spatially homogeneous,
$ {\rm {\bf H}}({\rm {\bf r}},\omega ) \to {\rm {\bf H}}(0,\omega )$.

For an asymmetric MN having, for example, an ellipsoid shape,
this allows us to write a circular electric field
\begin{equation}
 \label{eq4}
  [{\rm {\bf E}}_\text{rot} ({\rm {\bf r}},\omega )]_{x}
   \approx \ri{\frac{\omega}{c}}\left[ {\frac{H_{y}(0,
    \omega)}{c^2+a^2}}z - {\frac{{H_{x} (0,\omega )}}{a^2 + b^2} }y
     \right]a^{2} ,
\end{equation}
as in \cite{GT1}. The rest of the components of the rotational
field can be obtained from equation~(\ref{eq4}) by cyclic permutation of
indices. In equation~(\ref{eq4}), parameters $a$, $b$, and $c$ are the
ellipsoid half-axes in the directions $x$, $y$ and $z$, respectively.
The linear dependence of the rotational field on the coordinates
is easy to understand from equation~(\ref{eq2}), that defines this field.
If we make the time Fourier transform of equation~(\ref{eq2}), then the
right-hand side of this equation (as can be seen from equation~(\ref{eq3})
and the assumed inequality) can be considered as a constant not dependent
on coordinates. This means that $\rot{\rm{\bf E}}_\text{rot}({\rm {\bf r}},
\omega ) = \text{const}$. However, such an equality is possible only in the single
case when
\begin{equation}
 \label{eq5}
  [{\rm {\bf E}}_\text{rot} ({\rm {\bf r}},\omega )]_{j} = {\sum
   \limits_{k = 1}^{3}{\alpha _{jk} (\omega )\,x_{k}}}  \,,
\end{equation}
that is, the rotational field is linearly dependent on the coordinates.
Here, $\alpha $ is some matrix not dependent on ${\rm {\bf r}}$,
the components of which we will further specify.

From equation~(\ref{eq4}) one can see that the homogeneous external
magnetic field induces the coordinate-dependent rotational field
inside the particle. The internal field
${\rm {\bf E}}_\text{rot} ({\rm {\bf r}},t)$ generates the corresponding
density of current ${\rm {\bf j}}_\text{rot} ({\rm {\bf r}},t)$ in the MN.
As a result, the particle absorbs the energy from the EM field of the
incident laser wave.

\section{Dipole magnetic absorption}
\label{sec:2}

When the MN is illuminated by the field of a monochromatic EM wave
and the wave frequency is far from the plasmon resonance, the electrical
or the magnetic absorption can dominate depending on the
size of the MN \cite{G,TG}. However, if the size of the spherical
MN exceeds 50~\AA, then the magnetic absorption prevails the
electrical one for different polarizations of the incident light
(starting at frequencies smaller than the refraction frequency
of the electrons from the walls of the MN).

Relying on the Parseval correlation for the Fourier integral \cite{Br},
the energy of the magnetic absorption can be represented as
\begin{equation}
 \label{eq6}
  w_\text{m} = {\int\limits_{-\infty}^{\infty} W(t)\rd t}  =
   \frac{1}{2}\int\limits_{-\infty} ^{\infty} {\frac{{\rd\omega}}
    {2\piup} }\Re \int\limits_{V} {\rm {\bf j}}_\text{rot} ({\rm{\bf r}},
     \omega ){\rm {\bf E}}_\text{rot}^{\ast} ({\rm {\bf r}},\omega) \rd{\rm {\bf r}} ,
\end{equation}
where integration should be carried out over the entire
volume of the particle, $W(t)$ is the absorbed power.

The quantity ${\rm {\bf E}}_\text{rot} ({\rm {\bf r}},\omega )$ in accordance
with equation~(\ref{eq4}) is already known. Thus, the task remains to find
the Fourier component of the current density ${\rm {\bf j}}_\text{rot}
({\rm {\bf r}},\omega )$. In the general case, the current at the
point ${\rm {\bf r}}$ of a particle caused by the rotational field
${\rm {\bf E}}_\text{rot} ({\rm {\bf r}},\omega )$ can be written
as an integral over all electron velocities ${\bf v}$
\begin{equation}
 \label{eq7}
  {\rm {\bf j}}({\rm {\bf r}},\omega ) = 2e\left( {\frac{m}{2\piup
   \hbar}} \right)^{3}{\int_{ - \infty}^{\infty} {{\rm {\bf v}}
    f_{1}({\rm {\bf r}},{\rm {\bf v}},\omega )}\rd^{3}\upsilon} ,
\end{equation}
where $\upsilon=|{\bf v}|$, $f_{1} ({\rm {\bf r}},{\rm {\bf v}},\omega )$
is the Fourier component of a nonequilibrium distribution function, which
is considered as an addition to the equilibrium Fermi distribution function
$f_{0} ({\cal E})$ dependent only on the electron kinetic energy ${\cal E}$.
The function $f_{1} ({\rm {\bf r}},{\rm {\bf v}},\omega )$ is sought
as the solution of the corresponding linearized Boltzmann kinetic equation.
As a rule, it is written as a time dependent function (see, for example,
\cite{GT1}). If we perform the Fourier transform of this equation
and take into account the Fourier integral (\ref{eq3}) for the field,
and also
\begin{equation}
 \label{eq8}
  f_{1} ({\rm {\bf r}},{\rm {\bf v}},\omega ) = {\int\limits_{ -
   \infty}^{\infty}  {f_{1} ({\rm {\bf r}},{\rm {\bf v}},t)\,\re^{\ri \omega t}}}\rd t,
\end{equation}
we consequently get for $f_{1}$ the following equation
\begin{equation}
 \label{eq9}
  (\nu - \ri\omega )f_{1} ({\rm {\bf r}},{\rm {\bf v}},\omega ) + {
   \rm {\bf v}}{\frac{\partial f_{1} ({\rm {\bf r}},{\rm{\bf v}},
    \omega)} {\partial {\rm {\bf r}}}} + e{\rm {\bf E}}_\text{rot} ({\rm {\bf r}},
     \omega)\;{\frac{\partial f_{0} ({\cal E} )} {\partial {\rm {\bf p}}}} = 0,
\end{equation}
where $\nu $ is the frequency of bulk collisions of an electron,
${\rm {\bf p}} = m{\rm{\bf v}}$. The equation (\ref{eq9}) should
be supplemented by the corresponding boundary conditions. As such,
we have chosen the conditions for a diffuse reflection of electrons
from the internal walls of the particle
\begin{equation}
 \label{eq10}
  f_{1} ({\rm {\bf r}},{\rm {\bf v}},\omega )\vert_{S} = 0,
   \quad \quad\upsilon_{n} < 0.
\end{equation}

In equation~(\ref{eq10}), the value $\upsilon _{n} $ is the component
of the velocity normal to the surface $S$. The reasoning for
such boundary conditions can be found, in particular,
in \cite{KLY}.

To solve equation~(\ref{eq9}) with the boundary conditions (\ref{eq10}),
it is convenient to pass to the deformed variables
\begin{equation}
 \label{eq11}
  x_{i}' = R x_{i}/\rho_{i}\,,
   \quad\upsilon_{j}' = R \upsilon_{j}/\rho_{j}.
\end{equation}
The primed $x_{i}'$ and $\upsilon_{j}'$ symbols indicate the compression-stretching coordinates and velocities, which permit to modify an
ellipsoidal particle (with semiaxes $a, b, c$) into a spherical particle
[with the radius $R = (abc)^{1/3}$], and $\rho_x=a$, $\rho_y=b$, $\rho_z=c$.
Such modification changes only the shape of the particle leaving the
volume unchanged. Then the solution of equation~(\ref{eq9}) will take the form
\begin{eqnarray}
 \label{eq12}
  f_{1} ({\rm {\bf r}}',{\rm {\bf v}}',\omega ) = -
   \frac{e}{R^{2}}\frac{\partial f_{0}} {\partial{\cal E}}
    \sum\limits_{i,j = 1}^3 \alpha_{ij} (\omega )\;
     \upsilon'_{j}\; \rho_{j} \rho_{i}
      \left[ {x'_{i} + \upsilon'_{i} \frac{\partial}{\partial (
       \nu -\ri\omega )}} \right] \frac{1 - \re^{ - (\nu - \ri\omega )
        \,t'({\rm {\bf r}}',{\rm {\bf v}}')}}{\nu - \ri\omega}\,,
\end{eqnarray}
where
\begin{equation}
 \label{eq13}
  t'({\rm {\bf r}}',{\rm {\bf v}}') = {\frac{1}{\upsilon'^2}}\big[{
   \rm{\bf r}}'{\rm {\bf v}}' + \sqrt {(R^{2} - r'^{2})
    \upsilon'^{2} + ({\rm {\bf r}}'{\rm {\bf v}}')^{2}} \big].
\end{equation}
The diagonal components of the matrix $\alpha _{ij} $
are $\alpha _{jj} =0$, and the non-diagonal ones are expressed
through the corresponding components
of the magnetic field. For example,
\begin{equation}
 \label{eq14}
  \alpha _{xy} (\omega ) = - \ri{\frac{{\omega}} {c}}{
   \frac{a^{2}}{{a^{2} + b^{2}}} }H_{z} (0,\omega ).
\end{equation}
Two components of the matrix $\alpha _{ij}$ can be obtained by
cyclic permutation of the indices in equation~(\ref{eq14}). The remaining
three components can be found using skew-symmetry of $\alpha $,
that is, taking into account the property:
$\alpha_{xy}(\omega ) = - \alpha _{yx}(\omega )$.

Thus, to find the energy of the magnetic absorption of laser
pulses, it is necessary to perform the following steps: substitute
the found function $f_1({\bf r}, {\bf v},\omega )$ into equation~(\ref{eq7})
to obtain the Fourier component of the current density and then
substitute both the current density and the rotational field
given by equation~(\ref{eq4}) into equation~(\ref{eq6}).

The calculation of the absorption energy can be carried out avoiding
the described procedure using the Drude approach in the
study of the current.

\subsection{Drude approach}

Let the particle size be greater than the electron mean free path
within it. Then, we can express the rotational current in terms of
the rotational field as ${\rm {\bf j}}_\text{rot} (\omega ) =
\sigma_\text{m} (\omega )\;{\rm {\bf E}}_\text{rot} (\omega )$,
and equation~(\ref{eq6}) can be rewritten in the form
\begin{equation}
 \label{eq15}
  w_\text{m} = {\frac{1}{2}}{\sum\limits_{j = 1}^{3} \Re {{\int_{ -
   \infty}^{\infty}  {\sigma _\text{m} (\omega )\;{\frac{\rd\omega} {2
    \piup}}\;{\int_{V} {\vert {\rm {\bf E}}_\text{rot}^{j} ({\rm {\bf r}},
     \omega)\vert^{2}\rd{\rm\bf r}}}} }} } .
\end{equation}

Integration over the volume of the ellipsoid in equation~(\ref{eq15})
 can be replaced by integration over the sphere of an equivalent volume.

We will restrict ourselves further to the consideration of metal
nanoparticles with the spheroidal shape ($a = b \equiv\rho_{\bot}$,
$c \equiv\rho_{\|})$. Using equations~(\ref{eq4}) and elementary integration
over electron coordinates, the sum of integrals over a nanoparticle
volume will be
\begin{eqnarray}
 \label{eq16}
  \sum\limits_{j = 1}^{3} \int_{V} {\vert {\rm {\bf E}}_\text{rot}^{j}
   ({\rm {\bf r}},\omega )\vert}^{2} \rd{\rm {\bf r}}
    =
     \frac{V}{5}\left( \frac{\omega\rho_{\bot}}{c}\right)^{2}
      \left[ \frac{\rm{\bf H}_{\|}^{2}(0,\omega )}{2} +
       \xi\, {\rm {\bf H}}_{\bot}^{2} (0,\omega )
        \right].
\end{eqnarray}
Here,
\begin{equation}
 \label{eq17}
  \xi = \rho_{\|}^2/(\rho_{\|}^2 + \rho_{\bot}^2),
\end{equation}
${\rm {\bf H}}_{\|} ^{} (0,\omega )$ is the intensity of the
magnetic field along the spheroid rotation axis, and ${\rm {\bf H}}_{ \bot} (0,\omega )$ ---
across this axis. We note that the magnetic
absorption significantly depends on the magnetic field polarization.
Equation~(\ref{eq16}) can be generalized to the case of an
arbitrary coordinate system if the components
of the field are represented as:
\begin{equation}
 \label{eq18}
  {\rm {\bf H}}_{ \bot} ^{2} = {\rm {\bf H}}^{2} - {\rm
   {\bf H}}_{\|} ^{2} \,, \qquad {\rm {\bf H}}_{\|}
    = ({\rm {\bf H}}\cdot {\rm {\bf n}}){\rm {\bf n}}\,,
\end{equation}
where ${\rm {\bf n}}$ is a unit vector directed along the
spheroid axis of rotation. Then, with an account of
equations~(\ref{eq16})--(\ref{eq18}), the expression
(\ref{eq15}) transforms into
\begin{eqnarray}
 \label{eq19}
  w_\text{m} = \frac{V}{10}
   \left(\frac{\omega\rho_{\bot}}{c} \right)^{2} \xi \Re
    \int_{ - \infty} ^{\infty} {\frac{\rd\omega}{2\piup} }
     \sigma_\text{m} (\omega )\,
      \left\{ {\rm {\bf H}}^{2}(0,\omega ) - \frac{1}{2}
       \left( {1 - \frac{\rho_{\bot}^{2}} {\rho_{\|}^{2}} }
        \right)[{\rm {\bf H}}(0,\omega ){\rm {\bf n}}]^{2}
         \right\}.
\end{eqnarray}

The field ${\rm {\bf H}}(0,\omega )$ in equation~(\ref{eq19}) is given
by  equation~(\ref{eq3}), and the dependence $\sigma _\text{m} (\omega )$
for the metal particle in the Drude approach is
\begin{equation}
 \label{eq20}
  \sigma _\text{m} (\omega ) = {\frac{1} {4
   \piup}}{\frac{\omega_\text{pl}^{2}}{\nu - \ri\omega}}\,,
\end{equation}
where $\omega^2_\text{pl} = 4\piup n e^{2} / m$ is the frequency
of plasma oscillations of electrons in a metal.

Next, we will perform calculations for specific polarizations
of the magnetic wave relative to the MN orientation.
Consider the following two cases:

\smallskip

(i) The magnetic field is directed across the spheroid axis.
Then, the second term under the integral in equation~(\ref{eq19}) vanishes
and after substituting equations~(\ref{eq3}) and (\ref{eq20}) into
equation~(\ref{eq19}), we obtain an expression for the absorbed energy
\begin{equation}
 \label{eq21}
  w_{\text{m} \bot} = \frac{V\xi}{160}\frac{\nu}{\Gamma^2}
   \left( \frac{\omega_\text{pl} \rho_{\bot} } {c}\right)^{2}
    \vert{\rm {\bf H}}_{0} \vert ^{2}{
     \int\limits_{ - \infty}^{\infty} {\frac{\rd\omega} {2
      \piup} } {{\frac{{\omega ^{2}f(
       \omega )}}{\nu ^{2} + \omega^2}}}} \,,
\end{equation}
with
\begin{equation}
 \label{eq22}
  f(\omega ) = \left\{ {\exp {\left[ { - {\frac{{(\omega -
   \omega_\text{c} )^2}}{4\Gamma ^{2}}}} \right]} + \exp{\left[ { -
    {\frac{{(\omega + \omega_\text{c} )^2}}{4\Gamma^2}}} \right]}}
     \right\}^{2}.
\end{equation}

The exact integration in equation~(\ref{eq21}) is not possible in an
analytical form. However, taking into account that the frequency
of electron bulk collisions $\nu\approx 10^{13 }$~s$^{{\rm -}1} $
is a small value comparing to the frequencies $\omega\approx\omega_\text{c}
\sim 10^{15}$~s$^{{\rm -}1}$ which provide the main
contribution to the integral, one can neglect by $\nu$ in the
integrand of equation~(\ref{eq21}). Then, we come to twice the integral 
  $I_{\nu}  = {\int_{\nu} ^{\infty}  {f(
   \omega)}} \rd\omega.$

This integral can be calculated analytically even within pointed limits. However, it is not difficult to make sure that
the result of the integration is practically unchanged if the
lower limit of integration in the integral $I_{\nu}  $ is lowered
to zero. This means that the integral 
  $I = {\int_{0}^{\nu}
   {f(\omega )}} \rd\omega$
within the limits [0,$\nu $] is much smaller compared to $I_{\nu}$, 
and it can be omitted.
Thus, for an integral within the limits $[0,\infty ]$, we find:
\begin{eqnarray}
 \label{eq25}
  {\int\limits_{0}^{\infty}  {\left\{ {\exp {
   \left[ { -{\frac{(\omega - \omega_\text{c} )^2}{4\Gamma ^{2}}}} \right]} +
    \exp {\left[ { - {\frac{(\omega + \omega_\text{c} )^2}{4
     \Gamma^2}}} \right]}} \right\}^2}} \rd\omega =
      \sqrt {2\piup}  \;\Gamma\left[ {1 + \exp {\left( { - {
       \frac{{\omega_\text{c}^{2}}} {{2\Gamma^{2}}}}} \right)}} \right],
\end{eqnarray}
and as a result
\begin{equation}
 \label{eq26}
  w_{\text{m} \bot} = \frac{1}{80}{\frac{V\xi}{{\sqrt
   {2\piup}} } }{\frac{{\nu}}{{\Gamma}} }
    \left( \frac{\omega_\text{pl} \rho_{\bot} } {c}\right)^{2}
     \left( {1 + \re^{ - {\frac{\omega_\text{c}^{2} }{2
      \Gamma^{2}}}}} \right)\vert {\rm {\bf H}}_{0} \vert^{2},
\end{equation}
provided that $\omega\gg\nu $. If the ratio of $\omega_\text{c}/\Gamma>1$,
then the second term in parentheses can be omitted, but if the ratio
of $\omega_\text{c}/\Gamma\rightarrow 0$, then the result for $w_{\text{m} \bot}$ doubles.

\smallskip

(ii) Now, let us pass to the polarization of the EM wave, when
the magnetic field is directed along the spheroid axis. Then, the
scalar product in the second term under the integral in equation~(\ref{eq19})
is ${\rm {\bf H}}(0,\omega)\,{\rm {\bf n}} = \vert {\rm {\bf H}}
(0,\omega )\vert $ and substituting equation~(\ref{eq3}) into equation~(\ref{eq19}),
we find the expression
\begin{equation}
 \label{eq27}
  w_{\text{m}\|} = \frac{V}{80} \frac{\piup} {\Gamma^2}
   \frac{\rho_{\bot}^2 } {c^2} \vert {\rm {\bf H}}_{0}
    \vert^{2}\Re{\int\limits_{-\infty}^{\infty}\frac{\rd\omega}{2\piup}
     \omega ^{2}\sigma_\text{m}(\omega )f(\omega) }.
\end{equation}

Using equation~(\ref{eq20}) and the result of the integration
in equation~(\ref{eq25}), we arrive finally at the expression
for the energy absorbed with this polarization
\begin{equation}
 \label{eq28}
  w_{\text{m}\|} = {\frac{1}{160}}{\frac{V}{{\sqrt {2\piup} }}}{\frac{
   \nu}{\Gamma} }\left( {{\frac{{\omega _\text{pl} \rho_{ \bot} }}{c}}}
    \right)^{2}\left( {1 + \re^{ - {\frac{\omega_\text{c}^{2}} {2
     \Gamma^{2}}}}} \right)\vert {\rm {\bf H}}_{0} \vert^{2},
\end{equation}
provided that $\omega\gg\nu $.
Obviously, for a spherical MN $\xi=1/2$ in equation~(\ref{eq26})
and it is necessary to replace $\rho_{\bot}$ by $R$
in equations~(\ref{eq26}) and (\ref{eq28}).

If we assume that the monochromatic EM wave is falling
on the nanoparticle, that is the wave is described by equation~(\ref{eq1})
with $\Gamma \to 0$, then equations~(\ref{eq26}) and (\ref{eq28}) pass
to the known results from our previous calculations \cite{TG,GT1}
for the absorbed power $W_\text{m} = w_\text{m}\Gamma$ with an asymptotical
accuracy to the constant $\sqrt {{{\piup}  \mathord{\left/
{\vphantom {{\piup} {2}}} \right. \kern-\nulldelimiterspace} {2}}} /
2$. The accuracy is caused by the difference in the pulse form.

\subsection{Kinetic approach}

When the particle size is less than the electron mean
free path in it, the Drude approach cannot be used
anymore and it is necessary to replace it with the kinetic
approach. Note, that the latter also allows one to obtain the
correct results for the case when the electron mean free path
is less than the particle size. With the kinetic description of
the system, the current ${\rm {\bf j}}_\text{rot} (\omega )$ included
in equation~(\ref{eq6}), should be calculated by equation~(\ref{eq7}).
Using the found nonequilibrium distribution function given
by equation~(\ref{eq12}), we obtain
 \begin{align}
  \label{eq29}
   w_\text{m} &= \frac{2}{m}\frac{e^2}{R^4}\left(\frac{m}{{2\piup\hbar} }
    \right)^3 \Re{\sum\limits_{i j k l}^{3} {\rho_{i} \rho_{j}
     \rho_{k} \rho_l}}  {\int\limits_{ - \infty} ^{\infty}  {{\frac{{
      \alpha_{kj}^{\ast} (\omega )\alpha_{il} (\omega )}}{\nu - \ri
       \omega}}}} {\frac{\rd\omega}{2\piup} }\nonumber\\ &\times {\int\limits_{- \infty}^{
        \infty} {\upsilon'_{j} \upsilon'_{l}\delta (\upsilon^{2} -
         \upsilon _\text{F}^{2} ) \rd^{3}\upsilon}}{\int\limits_{V'} {x'_{k} x'_{i}
          \left[ {1 - \re^{ - (\nu - \ri\omega)t'(r',\upsilon ')}} \right]\rd r'}}.
\end{align}
Here, we have used a zero approximation with a small ratio of ${{k_\text{B} T}
\mathord{\left/ {\vphantom {{k_\text{B} T} {{\cal E}_\text{F}}} } \right.
\kern-\nulldelimiterspace} {{\cal E}_\text{F}}} $, when one can replace
the electron energy derivative by ${\partial f_{0}}/{\partial {\cal E}}
\to - \delta({\cal E}- {\cal E}_\text{F} )$, where ${\cal E}_\text{F} =
m\upsilon_\text{F}/2$ is the Fermi energy; $\upsilon_\text{F}$
is the electron velocity on the Fermi surface.
In equation~(\ref{eq29}), we omitted the term associated with the second
summand in the sum of equation~(\ref{eq12}). It is easy to show that the
contribution of this term tends to zero when integrating over all
electron coordinates, because it is even with respect to coordinates, whereas
the rotational field is an odd coordinate function. Then, the integration
over the coordinates in equation~(\ref{eq29}), associated with the first term
under the sum of equation~(\ref{eq12}) can be done exactly. Let us write the 
final result without details of the calculations that 
can be found in \cite{GT1},
\begin{eqnarray}
 \label{eq30}
  {\int_{V'} {x'_{k} x'_{i} \left[ {1 - \re^{ - (
   \nu - \ri\omega )\,t'(r',\upsilon ')}} \right]\rd r'}}
    = \piup \,R^{5}\left[\frac{\psi _1}{2}\left(
     \delta _{ki} - 3\frac{\upsilon'_{k}\upsilon'_{i}} {\upsilon'^2}
      \right) + 2\psi_{2} \frac{\upsilon'_{k} \upsilon'_{i}} {\upsilon'^{2}} \right].
\end{eqnarray}
Here,
\[
\psi_{1} (\upsilon ',\omega ) = {\frac{8}{15}} - {\frac{1}{\bar q}} +
{\frac{4}{{\bar q}^3}} - {\frac{24}{{{\bar q}^5}}} + \re^{ -
{\bar q}}{\frac{8}{{\bar q}^3}}\left( {1 + {\frac{3}{{\bar q}}} +
{\frac{3}{{\bar q}^2}}} \right),
\]
\[
\psi_{2} (\upsilon ',\omega ) = \frac{2}{5} - \frac{1}{\bar q} +
\frac{8}{3{\bar q}^2} - \frac{6}{{\bar q}^{3}} + \frac{32}{{\bar q}^5} +
\re^{ - {\bar q}} {\cal F}({\bar q}),
\]
\begin{equation}
 \label{eq31}
  {\bar q} = (\nu - \ri\omega ){\frac{2R}{\upsilon '}}\,,
   \qquad{\cal F}({\bar q}) = {\frac{2}{{\bar q}^2}}
    \left( {1 + {\frac{5}{\bar q}} + {\frac{16}{{\bar q}^2}} + {
     \frac{16}{{\bar q}^3}}} \right).
\end{equation}

The next integration in equation~(\ref{eq29}) over the velocity
space cannot be carried out in an analytical form.
It can be performed only if the frequency interval
($-\infty ,\infty $) is split symbolically into two parts
\begin{equation}
 \label{eq32}
  \int\limits_{ - \infty} ^{\infty}  \Phi (\omega
   )\,\rd\omega = 2{\int\limits_{0}^{\nu _\text{S}}  {\Phi_\text{LF} (\omega
    )\,\rd\omega}}   + 2{\int\limits_{\nu _\text{S}} ^{\infty}  {\Phi_\text{HF}
     (\omega )\,\rd\omega}}  ,
\end{equation}
one of which exceeds and the other one 
is less than the frequency ($\nu_\text{S}$) of electron
collisions with a particle surface
\begin{equation}
 \label{eq33}
  \nu_\text{S} = \upsilon_\text{F} / (2R).
   \end{equation}

The twos in equation~(\ref{eq32}) arose due to the fact that the
sub-integral expressions for $\Phi$ must be an even function
of $\omega $. The case with $\omega < \nu_\text{S} $ is hereinafter
referred to as low frequency (LF), and the case with $\omega > \nu_\text{S}$
--- to high frequency (HF). Let us consider first the last one.
It is based on using the approximations
\begin{equation}
 \label{eq34}
  \Re{\left\{ {{\frac{\psi_{1}(\upsilon',\omega )}{
   \nu - \ri\omega} }} \right\}}\approx{\frac{\upsilon'}{2R
    \omega^2}}\,, \qquad \Re{\left\{ {{\frac{\psi_{3}(\upsilon',
     \omega )}{{\nu -\ri\omega}} }} \right\}}
      \approx{\frac{\upsilon'}{8R\omega ^2}}\,,
\end{equation}
where $\psi_{3} = \psi_{2}-\frac{3}{4}\psi_{1}$.

Avoiding the repetition of the calculations, details of which can
be found in \cite{TG,GT1,GT2}, and taking into account equation~(\ref{eq19}),
we write the final result that can be obtained after integration
over all electron velocities
\begin{eqnarray}
\label{eq35}
 w_\text{m}^\text{HF} = \frac{9}{64}\frac{V\omega_\text{pl}^{2}}{4\piup \,
  c^{2}} \rho_{\bot} \upsilon_\text{F}\int\limits_{\nu_\text{S}}^{\infty}
   \left\{\eta_\text{H} (e_{s} )\;\xi^2
    \vert {\bf H}(0,\omega )\vert^{2} +
     \left[ \zeta_\text{H} (e_{s} ) - \eta_\text{H} (e_{s} )\;\xi^2
      \right]\vert {\bf H}(0,\omega )\, {\bf n}\vert^{2}
       \right\}\frac{\rd\omega}{2\piup}\, .
\end{eqnarray}

In equation~(\ref{eq35}), $\zeta_\text{H} (e_{s} )$ and $\eta_\text{H} (e_{s})$
are some functions that depend only on the spheroid eccentricity
$e_{s}$ (see their analytical forms in \cite{TG},
equations~(65) and (107), with changing $\zeta_\text{H}\to\rho_\text{H}$).
Graphically they are presented below in figure~\ref{fig1} depending
on the shape of MN, which is given by the ratio
$\rho_{\bot}/\rho_{\|}$ of the spheroid semiaxes.

From equation~(\ref{eq35}), we can see that the
frequency dependence of the integrand in the high-frequency
case (at arbitrary polarization) is contained only in the amplitude
of the magnetic field. Practically, this means that we can use the
result of the integration obtained above in the Drude approach.
If the lower integration limit in the Drude case was the value of $\nu$,
then this role here will be played by $\nu_\text{S}$. It is not difficult to
ensure that in the case with $\nu_\text{S} $ you can also diminish the lower
limit of integration to zero. Thus, using equations~(\ref{eq3}) and (\ref{eq25})
in equation~(\ref{eq35}), the energy of the magnetic component of
the laser EM wave with the polarization ($\bot$) (when the
magnetic field is directed across the spheroid axis)
can be obtained finally in the form
\begin{equation}
 \label{eq36}
  w_{\text{m}\bot}^\text{HF} = {\frac{9}{64}}{\frac{V\xi^2}{
   \sqrt{2\piup} }}{\frac{\omega_\text{pl}^{2}} {16\,c^{2}}}{
    \frac{\upsilon_\text{F}}{\Gamma}} \left( {1 + \re^{- {\frac{
     \omega_\text{c}^2}{2\Gamma^2}}}} \right)\eta_\text{H} (e_{s} )\,
      \rho_{\bot}\vert{\rm {\bf H}_{\bot}}\vert^{2},
\end{equation}
and with the ($\|$)-polarization, when, on the contrary,
the magnetic field is directed along this axis, as:
\begin{equation}
 \label{eq37}
  w_{\text{m}\|}^\text{HF} = {\frac{9}{64}}{\frac{V \rho_{\bot}}{{
   \sqrt {2\piup} }}}{\frac{{\omega_\text{pl}^{2}}} {16\,c^{2}}}{
    \frac{\upsilon_\text{F}} {\Gamma }}\left({1 + \re^{ - {\frac{{
     \omega_\text{c}^2}} {2\Gamma ^2}}}}\right)\zeta_\text{H} (e_{s})
      \vert {\rm {\bf H}_{\|}} \vert^{2}.
\end{equation}

Expressions (\ref{eq36}) and (\ref{eq37}) coincide for the
spherical particle, since $\eta_\text{H} (0) = 4\zeta_\text{H} (0)$.
Besides, if one tends $\Gamma\to 0$, then the expressions
(\ref{eq36}) and (\ref{eq37}) are transformed into the
absorption power $W_\text{m}=w_\text{m}\Gamma$, known from the previous
calculations \cite{TG,GT1} for the monochromatic
wave with the same accuracy mentioned above.

\begin{figure}[!t]
\centering\includegraphics[width=8.5cm]{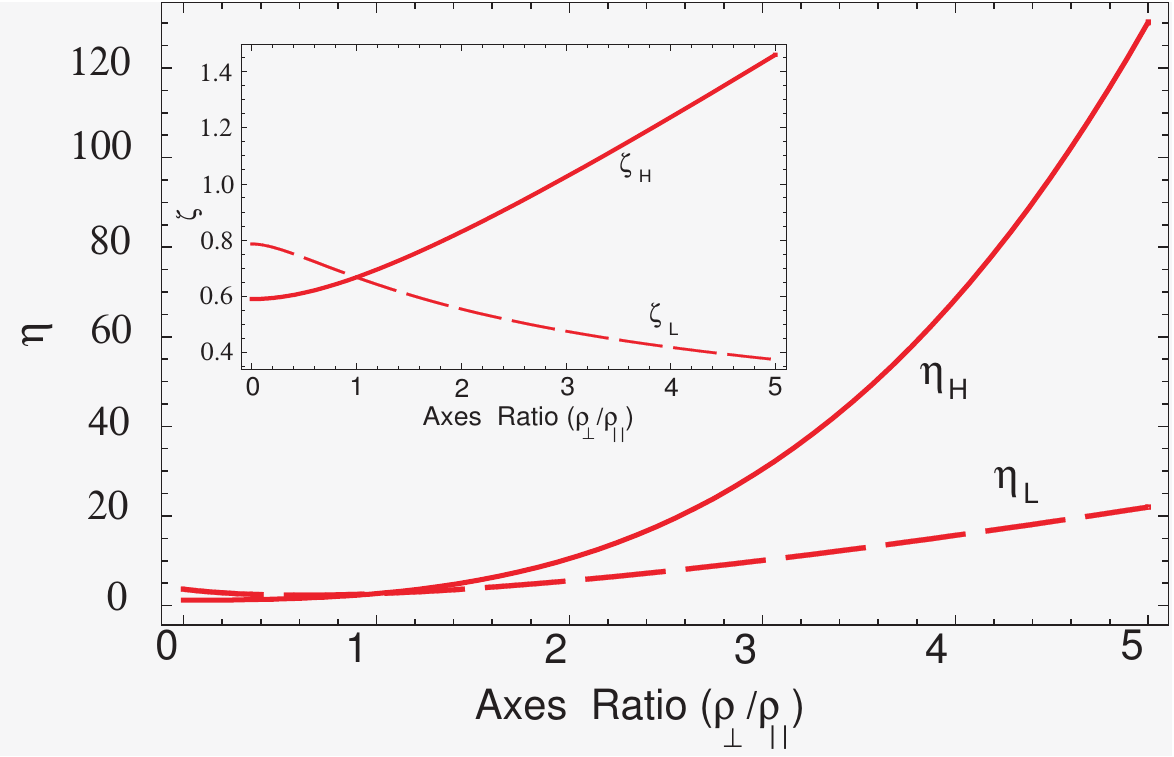}
\caption{(Colour online) Dependence of the function $\eta$ on the particle axes
ratio for the case of low $\omega < \upsilon_\text{F}/(2R)$ (dashed lines) and high $\omega > \upsilon_\text{F}/(2R)$ 
(solid lines) frequencies. In the inset: the same dependence
for the functions~$\zeta$ for the same frequency intervals.}
\label{fig1}
\end{figure}

It remains to consider the low frequency case. The next approximations
are applied in this case in the calculus of equation~(\ref{eq29})
\begin{equation}
 \label{eq38}
  \Re{\left\{ {{\frac{\psi_{1}
   (\upsilon ',\omega )}{{\nu - \ri\omega}} }} \right\}} \approx
    {\frac{R}{{3\,\upsilon'}}}\,, \qquad \Re{\left\{ {{\frac{\psi_{3}
     (\upsilon ',\omega )}{{\nu - \ri\omega}} }} \right\}} \approx
      {\frac{R}{{36\,\upsilon'}}}.
\end{equation}

The procedure of simple but cumbersome calculations
leads us to the next result
\begin{eqnarray}
 \label{eq39}
  w_\text{m}^\text{LF} = \frac{3}{32}V\frac{\omega_\text{pl}^2}
   {4\piup\,c^{2}}\frac{\rho_{\bot}^3} {\upsilon_\text{F}}\int
    \limits_{0}^{\nu_\text{S}}\left\{\eta_\text{L} (e_{s})\;\xi^2
     \vert {\bf H}(0,\omega)\vert^{2} +
      \left[\zeta_\text{L} (e_{s}) - \eta_\text{L} (e_{s})\;\xi^2
       \right]\vert {\bf H}(0,\omega)\,{\bf n}\vert^{2}
        \right\}\frac{\rd\omega} {2\piup}\,,
\end{eqnarray}
where $\zeta_\text{L} (e_{s} )$ and $\eta_\text{L}(e_{s})$ are some functions
dependent on the spheroid eccentricity $e_{s}$. Their analytical
form can be found in \cite{TG} (equations~(55) and (97),
with replacing $\zeta_\text{L}\to\rho_\text{L}$). In the inset of figure~\ref{fig1}, the behaviour of
these functions are shown as well. For a spherical particle, $\xi=1/2$,
$\zeta_\text{L}(0)=\zeta_\text{H}(0) = 2/3$, and $\eta_\text{L}(0) = \eta_\text{H}(0) = 8/3$.

The calculation of equation~(\ref{eq39})
needs the estimation of the integral
  $I_\text{L} = {\int_{0}^{\nu_\text{S}}
   {\omega^{2}f(\omega )}} \rd\omega$,
where $f(\omega)$ is given by equation~(\ref{eq22}).
As it was shown in \cite{GT2}, the input of this integral
is much less than the $I_{\nu_\text{S}}$ and one can neglect the first LF-term 
in the sum of equation~(\ref{eq32}) compared with the second one.
For an arbitrary angle $\theta $ between the direction of the
magnetic field and the spheroid rotation axis, we finally obtain
\begin{eqnarray}
 \label{eq41}
  w_\text{m} = w_\text{m}^\text{LF} + w_\text{m}^\text{HF} =
   \frac{9}{64}\frac{V}{\sqrt {2\piup}}\frac{\omega_\text{pl}^{2}}
    {16\,c^{2}}\frac{\upsilon_\text{F}}{\Gamma } \left(1 +
     \re^{-\frac{\omega_\text{c}^2} {2\Gamma^2}} \right)
      \rho_{\bot}\left[\zeta_\text{H} (e_{s} )
       \cos^{2}\theta + \eta_\text{H}(e_{s} )\,\xi^2
        \sin^{2}\theta \right]\vert {\bf H}_{ 0} \vert^{2},
\end{eqnarray}
where it is accounted that ${\bf H}_{\|}={\bf H}_{0}\cos\theta$
and ${\bf H}_{\bot}={\bf H}_{0}\sin\theta$.
For spherical MN, equation~(\ref{eq41}) is simplified to
\begin{eqnarray}
 \label{eq42}
  w_\text{m} =
   \frac{3}{64}\frac{V R}{\sqrt {2\piup}}\frac{\omega_\text{pl}^{2}}
    {8\,c^{2}}\frac{\upsilon_\text{F}}{\Gamma } \left(1 + \re^{ -
     \frac{\omega_\text{c}^2} {2\Gamma ^2}} \right)\,
      \vert {\bf H}_{ 0} \vert^{2}.
\end{eqnarray}

The results of equations~(\ref{eq26}) and (\ref{eq28}) obtained in
the Drude approach transform into the results for the spheroidal particle
given by equations~(\ref{eq36}) and (\ref{eq37}) or equation~(\ref{eq41}) obtained
in the kinetic approach, if we carry out a formal replacement
\begin{equation}
 \label{eq43}
  {\begin{array}{*{20}c}
   {\nu \to \displaystyle\frac{45}{64} \frac{\upsilon_\text{F}} {\rho_{\bot}}  \xi
    \eta_\text{H} 
     \qquad
\mbox{for}} \, \bot {\mbox{-polarization}}, \hfill \vspace{1ex}\\
 {\nu \to \displaystyle\frac{45}{32}\frac{\upsilon _\text{F}} {\rho_{ \bot}}
  \zeta_\text{H}
   \qquad \mbox{for}\, \|\mbox{-polarization}} \hfill \\
    \end{array}} 
\end{equation}
under a common condition $\omega\gg\nu $. Obviously, in the
case of a spherical particle, the above replacement will look like:
\begin{equation}
 \label{eq44}
  \nu \to {\frac{15}{16}}{\frac{\upsilon_\text{F}}
   {R}}\,,\quad \mbox{provided that}\  \omega\gg\nu .
\end{equation}

\section{Discussion of results}
\label{sec:5}

The energy of a magnetic absorption of a plane EM-wave by a spherical 
MN per unit time in a classical case can be written as \cite{LL}
\begin{equation}
 \label{eq45}
  w_\text{cl} = \frac{1}{80}\frac{V}{\piup}\omega\,\varepsilon''
   \left(\frac{\omega R}{c} \right)^2 {\vert{\bf H}_{0}\vert^2},
\end{equation}
where
\begin{equation}
 \label{eq46}
  \varepsilon'' = \frac{\nu}{\omega}
   \frac{\omega^2_\text{pl}}{\omega^2+\nu^2}.
\end{equation}
For the frequency domain of $\omega\gg\nu$,
equation~(\ref{eq45}) takes the form
\begin{equation}
 \label{eq47}
  w_\text{cl} = \frac{V}{80\piup}
   \left(\frac{\omega_\text{pl} R}{c}
    \right)^2 {\vert{\bf H}_{0}\vert^2},
\end{equation}
and the magnetic absorption of a plane EM-wave by a
spherical MN does not depend on the wave frequency.

Let us illustrate the previous analytical expressions
graphically. We calculate the ratio between the energy
of a magnetic field absorbed by the spheroidal MN from the laser pulses
per unit time, and the energy absorbed by the spherical MN
from the magnetic component of the plane EM-wave
\begin{equation}
 \label{eq48}
  S_\text{m} = \frac{w_\text{m}}{w_\text{cl}}.
\end{equation}
We limit ourselves only to the frequency interval
$\nu\ll\omega\ll\omega_\text{pl}$.

While studying the dependence of the optical properties of the MNs
on its shape, makes sense to compare the absorption in MNs
with different shapes, but equal in volume. We modify
the shape of a particle by changing the aspect ratio of the 
spheroid. The condition on the fixed volume of a particle
($V = {\frac{4\piup} {3}}\rho_{ \bot} ^{2} \rho_{\|} = \text{const}$)
with a given aspect ratio defines the
values of $\rho_{\bot}$ and $\rho_{\|}$. For example,
$\rho_{\bot} = R(\rho_{\bot}/\rho_{\|})^{1 / 3}$,
where $R$ is the radius of the sphere of an equivalent volume
\cite{G1}.

\begin{figure}[!t]
\centering\includegraphics[width=8.5cm]{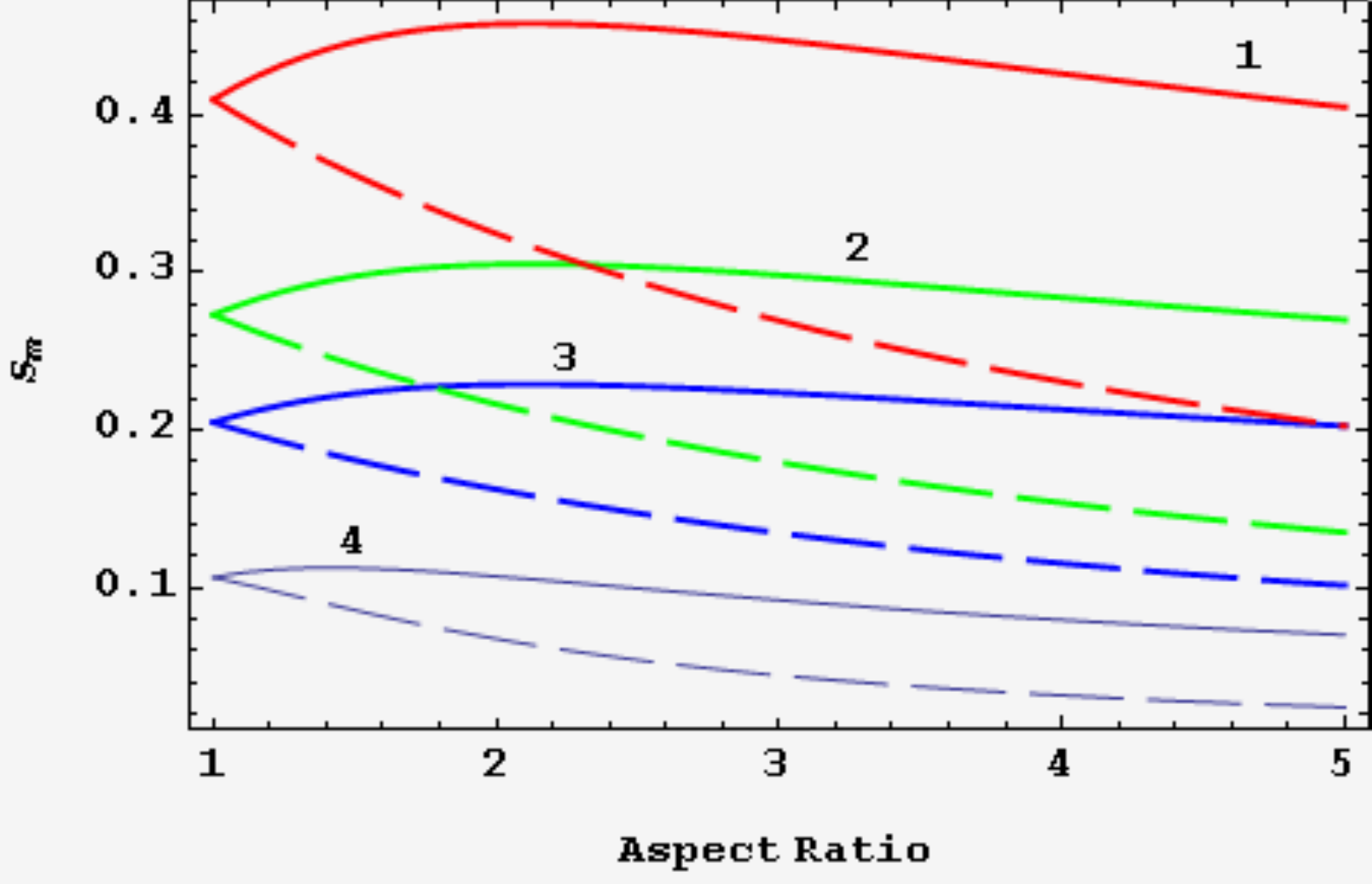}
\caption{(Colour online) 
Dependence of the energy absorbed by a spheroidal
metal nanoparticle with the polarization of the magnetic
field transverse ($ \bot $) to its axis of rotation on
the spheroid aspect ratio. Bold lines are for oblate MN
and dashed lines correspond to the prolate MN.
The curves are built for a particle whose volume
is equal to the volume of a spherical particle
with a radius $R = 100$~{\AA}, and for different~$\Gamma$: $2\cdot10^{14}$~s$^{ - 1}$ (curve 1),
$3\cdot 10^{14}$~s$^{- 1}$ (2), $4\cdot 10^{14}$~s$^{-1}$ (3).
The plot of the Drude dependence at $\Gamma = 2\cdot 10^{14}$~s$^{ - 1}$ is given by thin curves 4.}
\label{fig2}
\end{figure}

\begin{figure}[!t]
\centering\includegraphics[width=8.5cm]{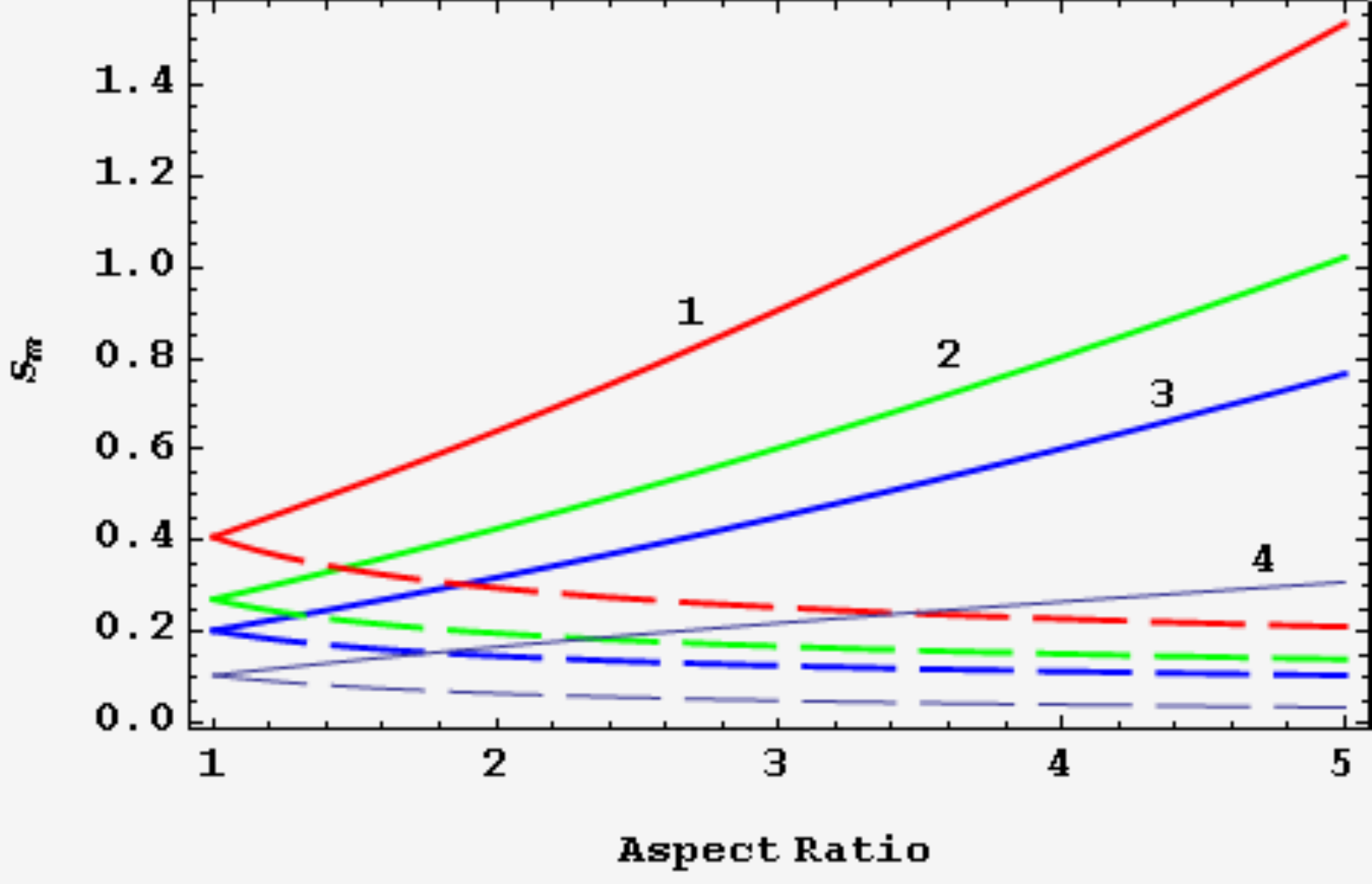}
\caption{(Colour online) The same as in figure~\ref{fig2} for $\|$-polarization of the magnetic field.}
\label{fig3}
\end{figure}

Below, we will study the effect of the MN shape on the
magnitude of the absorption energy of the magnetic component
of the laser field for two different polarizations of this field
(across and along the spheroid's rotation axis). The shape of
the particle will be determined by the spheroid aspect ratio.
In figures~\ref{fig2} and \ref{fig3}, the dependence of $S_\text{m}$ on the degree of
flattening or elongation of the spheroidal particle is depicted
for two polarizations of the magnetic field (at the frequency
of the surface plasmon $\omega_\text{c}= \omega_\text{pl} / \sqrt {3}$).

Calculations were carried out using
equations~(\ref{eq41}), (\ref{eq47}) and (\ref{eq48})
with parameters close to the Au \cite{K}:
$n=5.9\cdot10^{22}$~cm$^{-3}$,
$\upsilon_\text{F} \approx 1.39 \cdot 10^{8}$~cm/s,
$\omega_\text{pl}\approx 1.37 \cdot 10^{16}$~s$^{ -1}$,
and $\nu_{0^{\circ}\text{C}} = 3.39\cdot 10^{13}$~s$^{-1}$.
 Curves 1--3 correspond to the different duration of the
 incident pulse. The dependences of $S_\text{m}$ in the Drude approach
 constructed on the basis of equations~(\ref{eq26}) and
 (\ref{eq28}) are depicted by thin solid and dashed curves 4, respectively.
By comparing curves 1--3, we can see that the energy of laser
pulses with increasing duration is absorbed to a greater degree.

As one can see in figure~\ref{fig2} for the transverse polarization of the magnetic
field ($\theta=\piup/2$, $\bot$-polarization), the magnetic absorption
by the oblate MN increases, reaches a maximum at certain values of
the aspect ratio and then decreases. In the Drude case, this maximum
is achieved for particles of the elongated shape at $c/a = \sqrt {2}$,
and in the kinetic case --- at $c/a\approx 2$.

For the MN with a radius, for example,
100~{\AA}, one can make sure that the absorption in the
kinetic case is almost half the order of magnitude larger
(compare, e.g., curves 1 and 4 in figure~\ref{fig2}).

For the prolate MN in this polarization, the magnetic absorption
decreases only with the aspect ratio growth.

If we compare  the magnitudes of absorption for the spherical MN
that can be obtained in the Drude and kinetic approaches
[equations~(\ref{eq26}) and (\ref{eq36})] for the same pulse
duration, then we will find
\begin{equation}
 \label{eq49}
  {\left. {{\frac{w^\text{HF}_{\text{m} \bot}} {w_{\text{m} \bot}} } }
   \right|}_{\max} = {\frac{15}{16}}
    \frac{\upsilon_\text{F}}{\nu R}.
\end{equation}
From equation~(\ref{eq49}) it follows that this ratio increases
with a decrease in the radius of the particle and does not
depend on $\Gamma $.

In the polarization of the magnetic field along the spheroid axis
of rotation ($\theta=0$, $\|$-polarization), the magnitude of the
absorption increases (see figure~\ref{fig3}) for an increasingly flattened
MN and decreases with the growth of its elongation.

The reason for
this is the fact that the number of closed electron orbits with an
increasing flattening of the particle increases for $\|$- and
falls for $\bot$-polarization of the magnetic field.

In contrast to the electric absorption, with the deviation
of the carrier frequency from the frequency of the surface
plasmon in a spherical particle, the maximum in the dependence
of the absorbed energy on the aspect ratio is not displaced,
and the other maxima associated with the particle shape do not
appear. This fact takes place for both polarizations just as
in the Drude, and in the kinetic descriptions of the process.
However, depending on the duration of the incident pulse, there
is a certain specificity. With an increase of the pulse duration,
the absorption intensity increases at different carrier
frequencies in the both polarizations.

As follows from equations~(\ref{eq28}) and (\ref{eq42}), the
energy absorbed by the spherical MN slightly depends on the
ratio of $\Gamma/\omega_\text{c}$ in both the Drude and kinetic cases.
Note that the spatial dimension of the pulse $c / \Gamma $
must not exceed the length of the carrier wave in a vacuum:
$\Gamma/\omega_\text{c} \geqslant 1 / (2\piup ) \approx 0.16$.

Thus, the absorption in non-spherical particles essentially depends
on the polarization of the magnetic field relative to the particle.
The magnetic absorption in the spheroidal MN at $ \bot $-polarization of the field weakly increases (in comparison
with the spherical particles) in the flattened particles and falls 
in the elongated particles of the same volume.
In the Drude approach, the absorption obtained in this polarization
for both elongated and flattened particles is less than
the one for the spherical particle. It is the smallest
for particles with an elongated shape.

In the case of $\|$-polarization,  the
absorption markedly increases in the flattened and decreases
in the elongated MN in comparison with the spherical MN.
The same trends are observed in the Drude approach.

Thus, the spheroidal elongated or flattened MN in comparison
with spherical ones can absorb the energy of the magnetic field
of ultrashort laser pulses more or less intensively depending on
the orientation of the magnetic field relative to the spheroid
rotational axis.

A similar study for electrical absorption was carried out by
us in \cite{TG}. 

From equations~(\ref{eq26}) and (\ref{eq36}),
or equations~(\ref{eq28}), (\ref{eq37}), one can only see that the
energy of the magnetic absorption under the Drude description
grows as a square of $\rho_{\bot}$, whereas in the kinetic
description, the growth takes place linearly with $\rho_{\bot}$.

\section{Conclusions}
\label{sec:6}
	
The theory of the surface and shape influence on the dipole magnetic
absorption of ultraviolet laser pulses by metal nanoparticles
is developed for the cases when the electron mean free path
exceeds or is less than the particle size.

 The obtained analytical expressions allow one to study the
 magnetic absorption of pulses depending on their duration,
  shape of the particle, and  polarization of the magnetic
 field. They are suitable in a wide range of applications.
 The correspondence between the results obtained in the
 Drude and kinetic approaches is discussed.

 The absorption in non-spherical particles essentially depends
 on the polarization of the magnetic field relative to the particle.
Thus, for the spheroidal MN, with the direction of the magnetic field
 across the spheroid rotation axis,
 the absorption increases in the flattened and falls
 in the elongated MN in comparison with the particles of a spherical
 shape of the same volume. In the case of the orientation of the
 magnetic field along the axis of the particle rotation,
 the situation is more pronounced compared with the spherical MN:
 the absorption appreciably increases in the flattened
 and falls in the elongated MN.

It is established that the energy of laser pulses with an increasing
duration is absorbed to a greater extent. For any pulse duration,
one can reveal the following tendency: at the transverse ($\bot$)
polarization of the magnetic field, the absorption increases for the
increasingly \textit{flattened} MN, reaching the maximum at the
spheroid aspect ratio $a/c\approx 2$, and then decreases with
the greater aspect ratio. At the longitudinal polarization ($\|$)
of the magnetic field, the absorption by flattened MN only
increases with the flattening of the MN.

For particles of the \textit{elongated} shape with an increasing 
 elongation of the particle, the absorption decreases at both
the $\bot$- and the $\|$-polarization of the magnetic field.

These tendencies maintain for the Drude description
of the processes as well.

For the fixed pulse duration and with the both polarizations
($\|$ and $\bot$) of the magnetic field, the value of the magnetic
absorption obtained in the kinetic approach for particles with $R = 100$~{\AA} can be half an order of magnitude higher
than the corresponding values derived from the Drude approach.

Our results make it possible to control the effects of the
nanoparticle surface and the shape on the magnetic absorption of
ultrashort laser pulses by varying the particle shape, the
duration of a laser pulse and with changing its polarization.

\section*{Acknowledgements}
The work is supported by the Programme of the Fundamental Research of the Department of Physics and
Astronomy of National Academe of Science of Ukraine (NASU) (0116U002067).

\newpage

\ukrainianpart

\title{Вплив поверхні й форми наночастинки на дипольне 
магнітне поглинання ультракоротких лазерних імпульсів}
\author{М.І. Григорчук}
\address{
Інститут теоретичної фізики ім.~М.М.~Боголюбова НАН України, \\
 вул.~Метрологічна, 14-б, 03143 Київ, Україна
}

\makeukrtitle

\begin{abstract}
\tolerance=3000%
Розвинуто теорію поглинання енергії магнітного поля металевими наночастинками
несферичної форми, опромінених ультракороткими лазерними імпульсами різної 
тривалості. Вивчається вплив як поверхні частинки, так і її форми на величину 
поглинутої енергії. Для частинок сплюснутої чи витягнутої форми знайдена 
залежність цієї енергії від орієнтації магнітного поля відносно частинки, 
ступеня відхилення її форми від сферичної, тривалості імпульсу та від 
несучої частоти лазерного променя. Встановлений значний ріст у поглинанні, 
коли середня довжина вільного пробігу електрона збігається з розмірами 
частинки. Для обчислень використовувались підходи Друде та кінетичний 
і їх результати порівнювались.
\keywords металеві наночастинки, дипольне магнітне поглинання, 
ультракороткі лазерні імпульси, несферичні частинки
\end{abstract}

\end{document}